\documentclass[11pt,prd,a4paper,preprintnumbers,amsmath,amssymb,nofootinbib]{article}
\usepackage{feynmp-auto,expdlist}
\usepackage{amsmath,amsfonts,amssymb}
\usepackage{graphicx}
\usepackage{enumerate}
\usepackage{hyperref}
\usepackage{latexsym}
\usepackage{hepnicenames}
\usepackage{enumerate}
\usepackage{soul}
\usepackage[normalem]{ulem}
\usepackage{wasysym}
\usepackage{makecell}
\usepackage{bbm}

\font\tenrsfs=rsfs10 at 12pt
\font\sevenrsfs=rsfs7
\font\fiversfs=rsfs5
\newfam\rsfsfam
\textfont\rsfsfam=\tenrsfs
\scriptfont\rsfsfam=\sevenrsfs
\scriptscriptfont\rsfsfam=\fiversfs

\numberwithin{equation}{section}

\usepackage{mathrsfs}
\usepackage{braket}
\usepackage{titling}
\usepackage{amsmath}
\usepackage{slashed}
\usepackage{amssymb}
\usepackage{epsfig}
\usepackage{graphicx}
\usepackage{color}
\usepackage{rotating}
\usepackage{hyperref}
\usepackage[margin=1.in]{geometry}
\usepackage[table,xcdraw,dvipsnames]{xcolor}
\usepackage[compress,numbers,sort]{natbib}
\usepackage{colortbl}
\usepackage{pdflscape}
\usepackage{color}
\usepackage{mathtools}
\usepackage{colortbl}
\usepackage{comment}
\definecolor{Gray}{gray}{0.95}
\definecolor{RGray}{gray}{0.85}
\definecolor{CGray}{gray}{0.93}
\definecolor{piggypink}{rgb}{0.99, 0.87, 0.9}
\definecolor{babyblue}{rgb}{.67,.83,.99}

\definecolor{nicered}{rgb}{0.7,0.1,0.1}
\definecolor{nicegreen}{rgb}{0.1,0.5,0.1}
\definecolor{red}{rgb}{1.0, 0, 0}
\definecolor{niceblue}{rgb}{0,0,0.8}
\definecolor{red}{rgb}{1.0, 0, 0}
\hypersetup{colorlinks,bookmarksopen,bookmarksnumbered,
linkcolor=blus,pdfstartview=FitH,urlcolor=rossos,citecolor=verde}
\allowdisplaybreaks

\definecolor{rosso}{cmyk}{0,1,1,0.4}
\definecolor{rossos}{cmyk}{0,1,1,0.55}
\definecolor{rossoc}{cmyk}{0,1,1,0.2}
\definecolor{blu}{cmyk}{1,1,0,0.3}
\definecolor{blus}{cmyk}{1,1,0,0.6}
\definecolor{bluc}{cmyk}{1,1,0,0.1}
\definecolor{verde}{cmyk}{0.92,0,0.59,0.25}
\definecolor{verdec}{cmyk}{0.92,0,0.59,0.15}
\definecolor{verdes}{cmyk}{0.92,0,0.59,0.4}

\def\eq#1{{Eq.~(\ref{#1})}}

\def\fig#1{{Fig.~\ref{#1}}}

\def\sect#1{{Sect.~\ref{#1}}}

\def\app#1{{App.~\ref{#1}}}

\def\Tr{\mbox{Tr}\,}
\def\tr{\mbox{Tr}\,}

\def\diag{\mbox{diag}\,}

\renewcommand{\bar}{\overline}

\newcommand{\beq}{\begin{equation}}
\newcommand{\eeq}{\end{equation}}
\newcommand{\bea}{\begin{eqnarray}}
\newcommand{\eea}{\end{eqnarray}}

\renewcommand{\[}{\left[}

\def\be{\begin{equation}}
\def\ee{\end{equation}}

\usepackage[compat=1.1.0]{tikz-feynman}
\usepackage{graphicx}

\begin{document}

\begin{center}  
{\LARGE
\bf\color{blus} 
Flavour constraints on light spin-1 bosons \\
within a chiral Lagrangian approach 
} \\
\vspace{0.8cm}

{\bf Luca Di Luzio, Gabriele Levati, 
Paride Paradisi, Xavier Ponce D\'iaz}\\[7mm]

{\it Istituto Nazionale di Fisica Nucleare (INFN), Sezione di Padova, \\
Via F. Marzolo 8, 35131 Padova, Italy}\\[1mm]
{\it Dipartimento di Fisica e Astronomia `G.~Galilei', Universit\`a di Padova,
 \\ Via F. Marzolo 8, 35131 Padova, Italy
}\\[1mm]

\vspace{0.3cm}

\begin{quote}
We discuss the construction of the chiral Lagrangian for a light spin-1 boson, here denoted as $X$, featuring both vector and axial-vector couplings to light $u,d,s$ quarks. Focusing on $\Delta S = 1$ transitions, we show that there are model-independent tree-level contributions to $K^\pm \to \pi^\pm X$, sourced by Standard Model charged currents, which receive an 
$m^2_K / m_X^2$ 
enhancement from the emission of a longitudinally polarized $X$. This flavour observable sets the strongest to date model-independent bound on the diagonal axial-vector couplings of $X$ to $u,d,s$ quarks for $m_X < m_K - m_\pi$, superseding the bounds arising from beam-dump and collider searches.

\end{quote}

\thispagestyle{empty}
\end{center}

\bigskip

\tableofcontents

\clearpage

\section{Introduction}

The lack of new particles at the LHC may be suggestive of the fact 
that they are either too heavy to be 
directly
produced or too weakly coupled 
to Standard Model (SM) particles.
New Physics (NP) models containing new feebly interacting massive particles with sub-GeV masses are currently among the most studied 
NP scenarios both theoretically and experimentally.
Many of these studies were dedicated to the dark photon~\cite{Holdom:1985ag,Pospelov:2007mp}, 
a new massive spin-1 particle which is kinetically mixed with 
the ordinary photon and that could act as a portal to a dark sector.
Dark photon searches have been conducted 
by a number of experiments, including beam-dump~\cite{Bjorken:2009mm}, 
fixed-target~\cite{APEX:2011dww,Merkel:2014avp}, 
collider~\cite{BaBar:2009lbr,Curtin:2013fra,BaBar:2014zli,LHCb:2017trq,Anastasi:2015qla,LHCb:2019vmc,CMS:2019buh}, as well as meson decay~\cite{Bernardi:1985ny,KLOE-2:2011hhj,KLOE-2:2012lii,WASA-at-COSY:2013zom,HADES:2013nab,NA482:2015wmo,NA62:2019meo} experiments. 

Moreover, comprehensive analyses aiming at probing a light spin-1 boson $X$ with general couplings to quarks and leptons have been also carried out (see e.g.~\cite{Kahn:2016vjr,Ilten:2018crw,Baruch:2022esd,Asai:2022zxw}).
If $X$ is coupled to SM particles through a non-conserved current, 
such as the axial-vector current, processes which are enhanced by the ratio $({\rm energy}/m_X)^2$ involving the longitudinal mode
of $X$ are generally induced. The same happens if $X$ is coupled to 
a tree-level conserved current which is broken by the chiral anomaly \cite{Dror:2017ehi,Dror:2017nsg}.
These energy-enhanced contributions generally provide the dominant 
effects both to the production mechanisms of $X$ in high-energy
experiments, as well as to flavor-changing neutral current processes such as 
$K^\pm\to\pi^\pm X$. 

The aim of this paper is to revisit the sensitivity to the above NP scenarios of the rare decay $K^\pm\to\pi^\pm X$ induced by an underlying $s\to d$ quark transition.
In order to accomplish this task, we will extend previous studies by constructing the most general $\Delta S = 1$ chiral Lagrangian up to the order $\mathcal{O}(p^4)$, which will enable us to account for all of the dominant effects stemming from weak interactions. 
Indeed, the lowest-order $\mathcal{O}(p^2)$ terms of 
chiral perturbation theory ($\chi$PT) will capture the weak effects to the $s\to d X$ transition discussed in~\cite{Dror:2017ehi,Dror:2017nsg} and arising from the one-loop exchange of the $W$-boson and up-quarks.
Instead, as we will see, weak effects stemming from the 
$\Delta S=1$ four-quark Lagrangian~\cite{Buchalla:1995vs}, can be included only by keeping $\mathcal{O}(p^4)$ terms in $\chi$PT.
Although subleading in the chiral expansion, the latter contributions 
to $K^\pm\to\pi^\pm X$ arise already at tree level 
(they can be thought as arising from initial or final
state radiation of $X$ from the external quark legs of the 
$\Delta S=1$ effective Lagrangian) and therefore their inclusion appears to be mandatory.

Moreover, the tree-level weak contributions discussed in this work are
model-independent and therefore they represent a general and robust prediction of any ultraviolet (UV) complete NP model entailing a light spin-1 boson. 
Instead, the loop-induced effects discussed in~\cite{Dror:2017ehi,Dror:2017nsg} are sensitive to the specific UV completion responsible for the mass generation of $X$ (see e.g.~\cite{DiLuzio:2022ziu}).

The paper is organised as follows. In \sect{sec:2}, we will present the general derivation of the $\Delta S =1$ 
chiral Lagrangian, as well as the related Feynman rules for spin-1 bosons up to the $\mathcal{O}(p^4)$ order.
In \sect{sec:3}, we will compute the $K^\pm\to\pi^\pm X$ decay rate in $\chi$PT exploiting the
Feynman rules derived in \sect{sec:2}, comparing our tree-level effects with the results 
obtained at one-loop level in~\cite{Dror:2017ehi,Dror:2017nsg}. In \sect{sec:numerics}, we will discuss
our flavour bounds vs.~beam-dump and collider searches as reported in~\cite{Ilten:2018crw,Baruch:2022esd}.
\sect{sec:concl} is dedicated to our conclusions, 
while more technical details about the construction 
of the chiral Lagrangian are deferred to 
\app{appendix}.


\section{$\Delta S = 1$ chiral Lagrangian for spin-1 bosons}
\label{sec:2}

The most general Lagrangian describing the lowest-order
interactions of a new spin-1 particle $X$ with SM fermions includes both vectorial and axial couplings and, focusing on the interactions with the lightest quark flavours $q = (u,d,s)^T$, it can be written as
\begin{equation}
\label{eq:X_Lag_with_SM_Fermions}
\mathcal{L}^{\rm int}_X = g_x X_\mu \, \bar{q}\gamma^\mu (x_V + x_A \gamma_5) q \, , 
\end{equation}
where $g_x$ measures the strength of the universal coupling of $X$ to quarks. The vectorial and axial charges, $x_{V,A}$, are defined in 
flavour space and include 
off-diagonal entries in the 2-3 sector. 

\subsection{Lowest-order chiral Lagrangian}

At energies above few GeV, the Lagrangian of Eq.~\eqref{eq:X_Lag_with_SM_Fermions} can be directly employed to analyse the interactions of $X_\mu$ with quarks. Here, instead we focus on the low-energy range below the GeV scale, where we can resort 
to $\chi$PT -- see e.g.~\cite{Gasser:1984gg,Pich:1995bw}.
In order to construct our $\chi$PT in the presence of $X_\mu$, we proceed as follows. 
Let us consider the massless QCD Lagrangian with chiral symmetry group $G = SU(3)_L \times SU(3)_R$ 
\begin{equation}
\label{eq:Free_QCD_Lag}
\mathcal{L}_{\text{QCD}}^0 = 
-\frac{1}{4} G^a_{\mu\nu}G_a^{\mu\nu} 
+ i \bar{q}_L \gamma^\mu 
\left(\partial_\mu + ig_s\frac{\lambda_a}{2}A^a_\mu\right) q_L
+ i \bar{q}_R \gamma^\mu 
\left(\partial_\mu + ig_s\frac{\lambda_a}{2}A^a_\mu \right) q_R \, ,
\end{equation}  
where $q = (u, d, s)^T$ and $\lambda_a$ are the Gell-Mann matrices.

Chiral symmetry-breaking terms (like mass terms or interactions with external gauge fields other than gluons) can be implemented 
by introducing appropriate spurions ($a_\mu,\, v_\mu ,\, s, \,p$) as external source fields \cite{Gasser:1984gg}. 
Therefore, the resulting Lagrangian  
$\mathcal{L}_{\text{QCD}}^{\text{ext}}$ reads
\begin{equation}
\label{eq:QCD_Lag_With_Sources}
\begin{split}
\mathcal{L}_{\text{QCD}}^{\text{ext}} &= \mathcal{L}_{\text{QCD}}^0 + \bar{q} \gamma^\mu (v_\mu + a_\mu \gamma_5)q + \bar{q} (s - i p \gamma_5) q\\
&= \mathcal{L}_{\text{QCD}}^0 + \bar{q} \gamma^\mu (2 r_\mu P_R + 2 \ell_\mu P_L)q + \bar{q} (s - i p \gamma_5) q\,.
\end{split}
\end{equation}
where $2r_\mu = v_\mu + a_\mu$ and $2\ell_\mu = v_\mu - a_\mu$.
Its chiral counterpart is then found to be 
\begin{equation}
\label{eq:Chipt_Lag_With_Sources}
\mathcal{L}_{\chi \text{PT}}^{\text{ext}} = \frac{f_\pi^2}{4} \Tr{\left[D_\mu U^\dagger D^\mu U + U^\dagger \chi + \chi^\dagger U\right]} +\mathcal{O} (p^4)
\end{equation}
where $U (x) = \exp \left[ i \lambda_a \pi_a(x)/f_\pi \right]$ 
(with $f_\pi \simeq 92$ MeV) is the mesonic matrix transforming as $U(x) \rightarrow L U(x) R^\dagger$ under $SU(3)_L \times SU(3)_R$ and $\pi_a(x)$ are the Goldstone boson fields of $SU(3)_L \times SU(3)_R \to SU(3)_{V}$ spontaneous breaking. Moreover, we have defined
\begin{equation}
 D_\mu U = \partial_\mu U - i r_\mu U + i U \ell_\mu  \qquad \text{and} \qquad \chi = 2B_0 \, (s + i p)\,.
\end{equation}
In our model, described by the Lagrangian of \eq{eq:X_Lag_with_SM_Fermions}, 
the covariant derivative $D_\mu U$ reads
\begin{equation}
    D_\mu U= \partial_\mu U -i g_x X_\mu(Q^x_R U-U Q^x_L) \, , 
\end{equation}
where $Q^x_{R/L}=Q^x_V\pm Q^x_A$,  
while $Q^x_V=\diag{(x_V^u,\,x_V^d,\,x_V^s)}$ and
$Q^x_A=\diag{(x_A^u,\,x_A^d,\,x_A^s)}$. 

Expanding the Lagrangian in \eqref{eq:Chipt_Lag_With_Sources} and keeping only the lowest order terms in the NP coupling, we find  
%
\begin{align}
\label{eq:chiral_LO}
\mathcal{L}_{\chi \text{PT}}^{\text{ext}} \supset &-i g_x X_\mu (x_V^u-x_V^s) \left(\partial^\mu K^- K^+ -\partial^\mu K^+ K^- \right) -i X_\mu g_x(x_V^u-x_V^d) \left(\partial^\mu \pi^- \pi^+ -\partial^\mu \pi^+ \pi^- \right)\nonumber\\
    & + \left[-i g_x X_\mu x_V^{32}\,\left(\partial^\mu K^+ \pi^--\partial^\mu\pi^-K^+\right) +\text{h.c.}\right] \, , 
\end{align}
with the corresponding Feynman rules given in 
\fig{fig:FeynmanRuleLO}
(all momenta flow from left to right).

Note that all couplings in Eq.~\eqref{eq:chiral_LO} are of vector type. 
This is due to the fact that the matrix element of the axial-vector quark operators in Eq.~\eqref{eq:X_Lag_with_SM_Fermions} vanishes between external pseudo-scalar meson states. Moreover, in the limit of universal vector couplings, i.e. $x_V^u = x_V^d = x_V^s$, the $K^+K^- X$ and $\pi^+\pi^- X$ interaction terms vanish as well, as a result of the underlying $SU(3)_V$ chiral symmetry, while the $K^+\pi^- X$ vector coupling still survives as the flavour-changing current is not conserved. 
\begin{figure}[ht]
\centering
\begin{tikzpicture}[]
\begin{feynman}[]
\node [dot](a);
\vertex [left=1.5cm of a] (i1);
\vertex [right=1cm of a] (aux);
\vertex [above= 1cm of aux] (f1);
\vertex [below= 1cm of aux] (f2);
\vertex[right=1 cm of aux] (f3) {$= ig_x(p_1+p_2)_\mu(x_V^u-x_V^s)$};
\diagram*{
(i1) -- [charged scalar, edge label=$K^+(p_1)$] (a),
(a) -- [charged scalar,edge label'=$K^+(p_2)$] (f1),
(a) -- [boson, edge label'=$X_\mu$] (f2)
};
\end{feynman}
\end{tikzpicture}
\end{figure}
\begin{figure}[h]
\centering
\begin{tikzpicture}[]
\begin{feynman}[]
\node [dot](a);
\vertex [left=1.5cm of a] (i1);
\vertex [right=1cm of a] (aux);
\vertex [above= 1cm of aux] (f1);
\vertex [below= 1cm of aux] (f2);
\vertex[right=1 cm of aux] (f3) {$= ig_x(p_1+p_2)_\mu(x_V^u-x_V^d)$};
\diagram*{
(i1) -- [charged scalar, edge label=$\pi^+(p_1)$] (a),
(a) -- [charged scalar,edge label'=$\pi^+(p_2)$] (f1),
(a) -- [boson, edge label'=$X_\mu$] (f2)
};
\end{feynman}
\end{tikzpicture}
\end{figure}
\begin{figure}[!h]
\qquad\qquad\qquad\qquad\qquad
\begin{tikzpicture}[]
\begin{feynman}[]
\node [crossed dot](a);
\vertex [left=1.5cm of a] (i1);
\vertex [right=1cm of a] (aux);
\vertex [above= 1cm of aux] (f1);
\vertex [below= 1cm of aux] (f2);
\vertex[right= 1cm of aux] (f3){$=-i g_x x_V^{32}(p_1^\mu + p_2^\mu)$};
\diagram*{
(i1) -- [charged scalar, edge label=$K^+(p_1)$] (a),
(a) -- [charged scalar,edge label'=$\pi^+(p_2)$] (f1),
(a) -- [boson, edge label'=$X^\mu$] (f2)
};
\end{feynman}
\end{tikzpicture}
\caption{Feynman rules for the lowest-order chiral Lagrangian}
\label{fig:FeynmanRuleLO}
\end{figure}

Moreover, tree-level contributions to $\Delta S =1$ processes, such as 
$K^\pm\to\pi^\pm X$, are generated only if the couplings $x_V$ are flavor off-diagonal. Yet, even for flavour-diagonal couplings, irreducible flavour-violating effects to $x_V$ are loop-induced by the exchange of the $W$ boson and up-quarks (see e.g.~\cite{Dror:2017nsg}).
In the following, we will show that weak interactions provide 
additional sources of flavour-violation to the $\Delta S = 1$
chiral Lagrangian, already at tree level, when we include higher-order 
terms in the momentum 
expansion corresponding to four-quark operators.

\subsection{Chiral Lagrangian for weak interactions}

In the SM, at energies above the scale of chiral symmetry breaking, 
$\Delta S=1$ transitions are induced by the effective four-fermion 
Lagrangian~\cite{Buchalla:1995vs}
\begin{equation}
\label{eq:SMDelta1Lag}
    \mathcal{L}^{\Delta S= 1}_{\rm SM} = G \sum_{i=1}^{10} C_i(\mu) O_i (\mu) \qquad \text{with } \qquad G \equiv -\frac{G_F}{\sqrt{2}}V_{ud}V^*_{us} \, , 
\end{equation}
where 
%
\begin{equation}
\label{eq:DeltaS1Operators}
    \begin{array}{ll}
        Q_1 = 4(\bar{s}_L\gamma_\mu d_L) (\bar{u}_L \gamma_\mu u_L),  & \qquad \qquad Q_2 = 4(\bar{s}_L\gamma_\mu u_L) (\bar{u}_L \gamma_\mu d_L), \\
        Q_3 = 4(\bar{s}_L\gamma_\mu d_L) (\bar{q}_L \gamma_\mu q_L),  & \qquad \qquad Q_4 = 4(\bar{s}^\alpha_L\gamma_\mu d^\beta_L) (\bar{q}^\beta_L \gamma_\mu q^\alpha_L), \\
        Q_5 = 4(\bar{s}_L\gamma_\mu d_L) \sum_q(\bar{q}_R \gamma_\mu q_R),  & \qquad \qquad Q_6 = 4(\bar{s}^\alpha_L\gamma_\mu d^\beta_L) \sum_q(\bar{q}^\beta_R \gamma_\mu q^\alpha_R), \\
        Q_7 = 6(\bar{s}_L\gamma_\mu d_L) \sum_q e_q(\bar{q}_R \gamma_\mu q_R),  & \qquad \qquad Q_8 = 6(\bar{s}^\alpha_L\gamma_\mu d^\beta_L)\sum_qe_q (\bar{q}^\beta_R \gamma_\mu q^\alpha_R), \\
        Q_9 = 6(\bar{s}_L\gamma_\mu d_L) \sum_q e_q(\bar{q}_L \gamma_\mu q_L),  & \qquad \qquad Q_{10} = 6(\bar{s}^\alpha_L\gamma_\mu d^\beta_L)\sum_q e_q (\bar{q}^\beta_L \gamma_\mu q^\alpha_L),  \\
    \end{array}
\end{equation}
%
$q = {u,d,s}$, $e_u=2/3$ and $e_d = e_s = -1/3 $; $\alpha$ and $\beta$ are colour indices which, if unspecified, are assumed to be contracted between the two quarks in the same current.

%

The construction of the chiral counterpart to \eq{eq:DeltaS1Operators} proceeds in two steps: 
\begin{itemize}
\item In the first step, one constructs the chiral structures describing the product of two fermionic currents. These structures must possess the same chiral 
transformation properties of the corresponding quark currents and are 
obtained by exploiting the quark-hadron duality between the Lagrangians of 
Eqs.~\eqref{eq:QCD_Lag_With_Sources} and \eqref{eq:Chipt_Lag_With_Sources}. 
At low energies, one has  
\begin{equation}
\label{eq:QCD_Chipt_Duality}
 \int \mathcal{D}q\, \mathcal{D} \bar{q}\, \mathcal{D}G_\mu\, \exp \left( i \int d^4 x \,\mathcal{L}_{\text{QCD}}^{\text{ext}} \right) = \int \mathcal{D}U  \exp \left(i \int d^4 x \, \mathcal{L}_{\chi\text{PT}}^{\text{ext}}\right) \, , 
\end{equation}
and taking the functional derivatives of the QCD and the $\chi$PT actions with respect to the external sources one can readily find the chiral counterparts to the various Dirac structures. 
For instance, up to order $\mathcal{O} (p^2)$ one finds 
\begin{equation}
\label{eq:ChiralEquivalent}
    \begin{split}
        &\bar{q}^i_L \gamma^\mu q^j_{L} = \frac{\delta S_{\text{QCD}}}{\delta(\ell_\mu)_{ij}} \equiv \frac{\delta S_{\chi\text{PT}}}{\delta(\ell_\mu)_{ij}} =   \frac{i}{2} f_\pi^2 (D^\mu U^\dagger U)_{ji} = -\frac{1}{2} f_\pi^2 (L^\mu)_{ji} \, ,   \\
        &\bar{q}^i_R \gamma^\mu q^j_{R} = \frac{\delta S_{\text{QCD}}}{\delta(r_\mu)_{ij}} \equiv \frac{\delta S_{\chi\text{PT}}}{\delta(\ell_\mu)_{ij}} = \frac{i}{2} f_\pi^2 (D^\mu U U^\dagger)_{ji}  = -\frac{1}{2} f_\pi^2 (R^\mu)_{ji} \, , \\
        &\bar{q}^i_L q_{R}^j = -\frac{\delta S_{\text{QCD}}}{\delta(s-i p)_{ij}} \equiv -\frac{\delta S_{\chi\text{PT}}}{\delta(s- ip )_{ij}} = -\frac{B_0}{2} f_\pi^2 U_{ji}  \, , \\
        &\bar{q}^i_R q_{L}^j = -\frac{\delta S_{\text{QCD}}}{\delta(s+ip)_{ij}} \equiv -\frac{\delta S_{\chi\text{PT}}}{\delta(s + i p)_{ij}} = -\frac{B_0}{2} f_\pi^2 (U^\dagger)_{ji} \, ,
    \end{split}
\end{equation}
where in the previous expressions we have defined the chiral currents $L_\mu$ and $R_\mu$

\begin{equation}
    L_\mu \equiv i U^\dagger D_\mu U = -i D_\mu U^\dagger U \, ,  \qquad R_\mu \equiv i U D_\mu U^\dagger = -i D_\mu U U^\dagger\,.  
\end{equation}

 
\item In the second step, one decomposes the product of quark currents into irreducible representations of the flavour algebra by defining appropriate projectors. These are to be applied as well to the chiral realisation of the quark currents in order to obtain the desired operators in the chiral Lagrangian, classified according to the irreducible representation of the flavour algebra they belong to 
(see e.g.~\cite{Pich:2021yll, RBC:2001pmy, Lehner:2011fz}). 
Further details are discussed in \app{appendix}. 


\end{itemize}

After carrying out the program outlined above, 
we finally reproduce the $\Delta S = 1$ chiral 
Lagrangian of Ref.~\cite{Pich:2021yll}, 
which takes the following simple form 
\begin{equation}
    \label{eq:ChiptWeakSM}
\begin{split}
    \mathcal{L}^{\Delta S= 1}_{\text{eff}}=G f_\pi^4 &\left\{ g_{27} \left(L^3_{\mu,\,2}L^{\mu,\,1}_1+\frac{2}{3}L^1_{\mu,\,2}L^{\mu,\,3}_1-\frac{1}{3}L^3_{\mu,\,2}\tr{\left[L^\mu\right]})\right)+  g_8^S \, L^3_{\mu,\,2}\tr{\left[L^\mu\right]} \right.\\ 
    &+  \left. g_8\left(\tr{\left[\lambda L_\mu L^\mu\right]}+e^2 g_{\text{ew}}f_\pi^2\tr\left[\lambda U^\dagger Q U\right]\right)\right\} \, , 
\end{split}
\end{equation}
where $\lambda \equiv \frac{1}{2} (\lambda_6-i\lambda_7)$ is responsible for the $s\rightarrow d$ 
flavour transition and we have specialised 
$Q=\frac{1}{3} \text{diag}(2, -1, -1)$ 
to be the charge matrix for quarks.
Out of the pieces making up Eq.~\eqref{eq:ChiptWeakSM}, the first one transforms in the $(27_L, 1_R)$ representation of the flavour group, while the second and the third ones transform in the $(8_L, 1_R)$ and $(8_L, 8_R)$
representation, respectively.
Clearly, no singlet term can have any effect on $\Delta S = 1$ transitions.
The coefficients $g_{27}$, $g_8$, $g_8^S$ and $g_{\text{ew}}$ are functions of non-perturbative effective parameters, as well as of the Wilson coefficients of the weak operators, see Eq.~\eqref{eq:SMDelta1Lag}. Their values are found to be

\begin{eqnarray}
    &g_8=3.07\pm 0.14 \qquad \,\,\,\,\, \text{\cite{Pich:2021yll}\,,} \\
    &g_8^S=-1.17\pm 0.37 \qquad \text{\cite{Gerard:2005yk}\,,}\\
    &g_{27}=0.29\pm 0.02 \qquad \,\,\,\text{\cite{Pich:2021yll}\,,}\\
    &g_{\text{ew}}=-1.0\pm 0.3 \qquad \,\,\,\text{\cite{Ecker:2000zr}\,.}
\end{eqnarray}

Expanding the Lagrangian of Eq.~\eqref{eq:ChiptWeakSM} and keeping only 
the contributions relevant for our analysis, we find 
\begin{equation}
\begin{split}
    \label{eq:ChiPT27}
    \mathcal{L}^{\Delta S= 1}_{\text{eff}} 
    \supset & \,\,\frac{2}{3}f^2 g_{27} G \left(2\partial^\mu K^+ \partial_\mu \pi^- + g_x X_\mu\left[i\partial^\mu K^+ \pi^- (4x_A^u-x_A^d-3x_A^s+2x_V^u-2x_V^d)\right.\right.  \\ 
    &\left. \left. -i\partial^\mu\pi^- K^+ (4x_A^u-3x_A^d- x_A^s+2x_V^u -2x_V^d)+\text{h.c.}\right] \right)\\
    &+ 2f^2 g_8^S G g_x \, (x_A^u+x_A^d+x_A^s)\, X_\mu  \left[i \,\left(\partial^\mu K^+ \pi^--\partial^\mu\pi^-K^+\right) +\text{h.c.}\right] \\
    &+ 2f^2 g_8 G \left(\partial^\mu K^+ \partial_\mu \pi^- + g_x X_\mu\left[i\partial^\mu K^+ \pi^- (x_A^u+x_A^s+x_V^u-x_V^d)\right.\right.  \\
    &\left. \left. -i\partial^\mu\pi^- K^+ (x_A^u+x_A^d+x_V^u-x_V^s)+\text{h.c.}\right] \right) +
    2 f^4 G e^2 g_8 g_{\text{ew}} K^+ \pi^- \, , 
\end{split}
\end{equation}
which includes both a $K\pi$ mixing term as well as a flavour-violating 
$K^\pm\to\pi^\pm X$ interaction, as depicted in the Feynman rules of \fig{fig:FeynmanRuleWeak}.  
\begin{figure}[h]
\label{fig:FeynmanRuleMixing}
~~
\begin{tikzpicture}[]
\begin{feynman}[]
\node [crossed dot](a);
\vertex [left=1.5cm of a] (i1);
\vertex [right= 1.5cm of a] (f1);
\vertex[right=0.5 cm of f1] (f2) {~~$= i2f^2 G\left[ g_8(f^2  e ^2 g_{\text{ew}}+p^2)+\frac{2}{3}g_{27}p^2\right]$};
\diagram*{
(i1) -- [charged scalar, edge label=$K^+(p)$] (a),
(a) -- [charged scalar,edge label=$\pi^+(p)$] (f1),
};
\end{feynman}
\end{tikzpicture}
\end{figure}
\begin{figure}[!h]
\centering
\begin{tikzpicture}[]
\begin{feynman}[]
\node [crossed dot](a);
\vertex [left=1.5cm of a] (i1);
\vertex [right=1cm of a] (aux);
\vertex [above= 1cm of aux] (f1);
\vertex [below= 1cm of aux] (f2);
\vertex[right=7 cm of aux] (f3);
\vertex[above=0.05cm of f3] (f4) {$= ig_x2f^2G\left[g_8(p^\mu_1(x_A^u+x_A^s+x_V^u-x_V^d)+p^\mu_2(x_A^u+x_A^d+x_V^u-x_V^s))\right.$};
\vertex[below=0.05cm of f3] (f5) {$ \left.+ g^S_8(p^\mu_1+ p^\mu_2)(x_A^u + x_A^d+x_A^s) +\frac{g_{27}}{3}(p^\mu_1(4x_A^u-3x_A^d- x_A^s+2x_V^u\right.$} ;
\vertex[below= 0.8 cm of f3] (f6) {$ \left.-2x_V^d) +p^\mu_2(4x_A^u-x_A^d-3x_A^s+2x_V^u-2x_V^s))\right]$};
\diagram*{
(i1) -- [charged scalar, edge label=$K^+(p_1)$] (a),
(a) -- [charged scalar,edge label'=$\pi^+(p_2)$] (f1),
(a) -- [boson, edge label'=$X^\mu$] (f2)
};
\end{feynman} 
\end{tikzpicture}
\caption{Feynman rules for the chiral Lagrangian of weak interactions.}
\label{fig:FeynmanRuleWeak}
\end{figure}

Note that, if Eq.~\eqref{eq:X_Lag_with_SM_Fermions} contains an explicit
source of flavour violation, the latter Feynman rule has to be supplemented by the last contribution displayed in Fig.~\ref{fig:FeynmanRuleLO}.

Differently from the leading-order chiral Lagrangian of Eq.~\eqref{eq:chiral_LO}, we are now sensitive both to $x^f_V$ and $x^f_A$
couplings. Indeed, the hadronic matrix element $\langle K|O_{i}|\pi\rangle$, where $O_{i}$ are the weak operators defined in Eq.~\eqref{eq:DeltaS1Operators},
receives contributions from both vector and axial-vector currents, 
as there are not symmetry arguments to forbid them.
Again, for universal vector couplings, the $K^+\pi^- X$ 
interaction vanishes because of the underlying 
$SU(3)_V$ chiral symmetry.

\section{$K^\pm\to\pi^\pm X$ in $\chi$PT}
\label{sec:3}

In this section we will compute the decay rate of the process $K^\pm\to\pi^\pm X$ in $\chi$PT.
Exploiting the Feynman rules derived in the previous section,
we will first analyse the tree-level contributions 
and then we will compare them with 
the results obtained at one-loop level in~\cite{Dror:2017ehi}.

\subsection{Tree-level contribution}
\label{sec:tree}

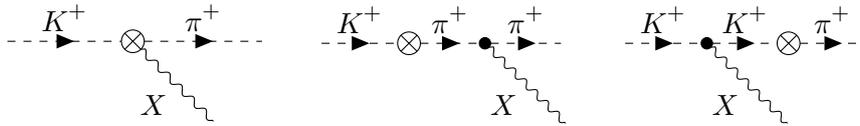
\begin{figure}[ht]
\centering
\begin{tikzpicture}[]
\begin{feynman}[]
\vertex(a);
\node[crossed dot,right = 1.5 cm of a](b);
\vertex[below left= 0.7 cm of b](n);
\vertex[right = 1.7 cm  of b](c);
\vertex[below right = of b](d);
\diagram*{
(a) -- [charged scalar, edge label=$K^+$] (b),
(d) -- [boson, edge label=$X$] (b) ,
(b) -- [charged scalar, edge label=$\pi^+$] (c),
};
\end{feynman}
\end{tikzpicture}\qquad
\begin{tikzpicture}[]
\begin{feynman}[]
\vertex(a);
\node[crossed dot,right = 1 cm of a](b);
\vertex[below left= 0.7 cm of b](n);
\node[dot, right = 1 cm of b](c);
\vertex[right = 1 cm  of c](e);
\vertex[below right = of c](d);
\diagram*{
(a) -- [charged scalar, edge label=$K^+$] (b),
(b) -- [charged scalar, edge label=$\pi^+$] (c),
(d) -- [boson, edge label=$X$] (c) ,
(c) -- [charged scalar, edge label=$\pi^+$] (e),
};
\end{feynman}
\end{tikzpicture}\qquad
\begin{tikzpicture}[]
\begin{feynman}[]
\vertex(a);
\node[crossed dot,right = 1 cm of b](c);
\vertex[below right= 0.7 cm of c](n);
\node[dot, right = 1 cm of a](b);
\vertex[right = 1 cm of c](e);
\vertex[below right = of b](d);
\diagram*{
(a) -- [charged scalar, edge label=$K^+$] (b),
(b) -- [charged scalar, edge label=$K^+$] (c),
(d) -- [boson, edge label=$X$] (b) ,
(c) -- [charged scalar, edge label=$\pi^+$] (e),
};
\end{feynman}
\end{tikzpicture}
\caption{Diagrams generating the tree-level transition $K^\pm\to\pi^\pm X$ in $\chi$PT.}
\label{fig:FeynmanDiagrams}
\end{figure}

At the tree level, the process $K^\pm\rightarrow\pi^\pm X$ 
is generated by the diagrams in \fig{fig:FeynmanDiagrams}.
The $X$ boson can be emitted either at the same vertex where the flavour transition takes place (first diagram) or at a different one. In the latter case (second and third diagrams) weak interactions prompt a flavour transition while the leg emission of an $X$ boson occurs at a different interaction point.



The total amplitude $\mathcal{M} = \mathcal{M}_8 + \mathcal{M}_8^S +
\mathcal{M}_{27} + \mathcal{M}_{\rm{ew}}$ receives four independent contributions proportional to $g_8,\,g_8^S,\,g_{27}$ and $g_{\rm{ew}}$ which are given by
\begin{align}
\label{eq:Canc8}
        \mathcal{M}_8
        &= 2 f_\pi^2 g_8 G g_x \varepsilon_\mu^*(q) \bigg[ p^\mu_1 \bigg(x_A^u + x_A^s + \frac{m_\pi^2}{m_K^2-m_\pi^2} (x_V^d-x_V^s)\bigg) 
        \nonumber\\ & \qquad \qquad \qquad \qquad ~
        + p^\mu_2 \bigg(x_A^u + x_A^d + \frac{m_K^2}{m_K^2-m_\pi^2} (x_V^d-x_V^s)\bigg) \bigg] \, , 
        \\
        \mathcal{M}^S_8 
        &= 2 f_\pi^2 g_8^S G g_x \varepsilon_\mu^*(q) (p_1 + p_2)^\mu (x_A^u + x_A^d+ x_A^s)
        \, , \\
        \mathcal{M}_{27}        
        &= \frac{2f^2_\pi g_{27} G g_x}{3}\, \varepsilon_\mu^*(q) \bigg[ p^\mu_1 \bigg( 4 x_A^u - 3 x_A^d- x_A^s + 2\frac{m_\pi^2}{m_K^2-m_\pi^2} (x_V^d-x_V^s)\bigg)\nonumber\\
        & \qquad \qquad \qquad  + p^\mu_2 \bigg( 4 x_A^u -  x_A^d - 3 x_A^s+ 2 \frac{m_K^2}{m_K^2-m_\pi^2} (x_V^d-x_V^s)\bigg) \bigg] \, , 
        \\
    \mathcal{M}_{\text{ew}} 
    & = - 2 f_\pi^2 e^2 g_8 g_{\text{ew}} G g_x (p_1+p_2)^\mu \varepsilon^*_\mu(q)  \frac{f_\pi^2}{m_K^2-m_\pi^2} (x_V^s-x_V^d)\,.
\end{align}
On the other hand, the decay rate can be written as
\begin{align}
\begin{split}
    \Gamma = \frac{1}{2 m_K}
    \frac{\left\vert \bar{\mathcal{M}}\right\vert^2}{8\pi}
    \left[1-2\left(\frac{m_X^2+m_{\pi}^2}{m_K^2}\right)+\left(\frac{m_X^2-m_{\pi}^2}{m_K^2}\right)^2\right]^{1/2} \, , 
\end{split}
\end{align}
where the total unpolarised amplitude squared is given by
\begin{equation}
    \begin{split}
    \label{amplitude_squared}
            |\bar{\mathcal{M}}|^2 &= \frac{g_x^2}{ m_X^2 \left(m_K^2-m_\pi^2\right)^2} [(m_K-m_{\pi})^2-m_X^2][(m_K+m_{\pi})^2-m_X^2] \bigg[x_V^{32} (m_K^2-m_\pi^2)\\ 
            & - 2 g_{\text{ew}} e^2 f^4 G g_8 (x^s_V-x^d_V) - f^2 G  \bigg( g_8^S (m_K^2-m_\pi^2) (x_A^u + x_A^d+x_A^s) \\
            &  + \frac{2}{3}g_{27} m_K^2 (x^s_V-x^d_V+2x^d_A+2x^s_A-4 x^u_A) + g_8 m_\pi^2 (x^s_V-x^d_V+x^d_A+x^s_A+2 x^u_A)\\
            &  - \frac{2}{3}g_{27} m_\pi^2 (x^s_V-x^d_V -2x^d_A- 2 x^s_A+4 x^u_A) - g_8 m_K^2 (x^s_V-x^d_V - x^d_A-x^s_A-2 x^u_A) \bigg)\bigg]^2 \, . 
    \end{split}
\end{equation}
Assuming generation universality of the couplings, i.e. $x_{V,A}^u = x_{V,A}^d = x_{V,A}^s$, and taking the limit $m_K \gg m_X, m_{\pi}$, one can find the simple expression
\begin{align}
\label{eq:gamma_simplified}
\begin{split}
    \Gamma \approx \frac{m_K}{2\pi} \left(\frac{m_K}{m_X}\right)^2 G_F^2 f_{\pi}^4 \, |V_{us}|^2 \, g_x^2 \, (x_A^u)^2 
    \left(g_8 + \frac{3}{4} g^S_8 \right)^2\, .
\end{split}
\end{align}
A few important comments are in order:
\begin{itemize}
    \item In the limit of universal vector couplings, the decay rate of 
    $K^\pm\to\pi^\pm X$ becomes independent of these couplings as a result of the underlying $SU(3)_V$ chiral symmetry.
    \item The enhancement factor $(m_K/m_X)^2$ in Eq.~\eqref{eq:gamma_simplified} for small $m_X$ is conceptually similar to the enhancement obtained in \cite{Dror:2017ehi,Dror:2017nsg}. This enhancement here is produced by the longitudinal component of the polarization vector: $\sum\varepsilon_\mu^*(q)\varepsilon_\nu(q)=-\eta_{\mu\nu}+\frac{q_\mu q_\nu}{m_X^2}$. 
\end{itemize}

In order to see where we stand, we write the branching ratio of 
$K^\pm\to\pi^\pm X$ as
\begin{align}
\label{eq:numerics}
    \mathcal{B}(K^+\to \pi^+X) \approx \frac{\Gamma(K^+\to \pi^+X)}{\Gamma(K^+\to \mu^+ \nu)} \times \mathcal{B}(K^+\to \mu^+\nu) \, ,
\end{align}
where $\Gamma(K\to \mu\nu)\approx m_K m^2_{\mu}|V_{us}|^2 f^2_K G^2_F/4\pi$ 
and ${\mathcal{B}}(K\to \mu\nu)\approx 64\%$. Moreover, 
we assume the equality $f_K = f_{\pi}$ which holds
in the $SU(3)$ chiral limit. Finally, exploiting the E949 measurement $\mathcal{B}(K^+\to \pi^+\nu\nu)=( 1.73^{+1.15}_{-1.05})\times 10^{-10}$~\cite{E949:2008btt}, we obtain the $2\sigma$ level constraint $\text{BR}(K^+\to \pi^+X)\lesssim 4\times 10^{-10}$. As a result, we find the following bound
\begin{align}
g_x \, x_A^u \lesssim 3\times 10^{-6}
\left(\frac{m_X}{0.1\,{\rm{GeV}}}\right)\,,
\end{align}
where the charges $x_A^u$ and $x_A^d$ are typically expected to be of order one. The above result will be fully confirmed by the numerical analysis of \sect{sec:numerics}.

\subsection{One-loop contribution}

\begin{figure}[h]
    \centering
    \begin{tikzpicture}
            \label{fig:loopA}
  \begin{feynman}
    \vertex (a) {$s$};
    \vertex [right=2cm of a] (b);
    \vertex [above right=2 cm of b] (f1) {$d$};
    \vertex [below right=2 cm of b] (c);
    \vertex [above right=1 cm of b] (d);
    \vertex [right=1 cm of a] (e);
    \diagram* {
      (a) -- [fermion] (e) -- [fermion, edge label'= $u_i$] (b) -- [fermion, edge label'= $u_i$] (d) -- [fermion] (f1),
      (b) -- [boson, edge label'=$X$] (c),
      (e) -- [boson, quarter left, edge label=$W$] (d),
    };
  \end{feynman}
\end{tikzpicture}
    \begin{tikzpicture}
            \label{fig:loopB}
  \begin{feynman}[]
    \vertex (a) {$s$};
    \vertex [right=2cm of a] (b);
    \vertex [above right=2 cm of b] (f1) {$d$};
    \vertex [below right=2 cm of b] (c);
    \vertex [above right=1.5 cm of b] (d);
    \vertex [above right=0.5 cm of b] (e);
    \diagram* {
      (a) -- [fermion] (b) -- [fermion, edge label'= $s$] (e) -- [fermion, edge label'= $u_i$] (d) -- [fermion] (f1),
      (b) -- [boson, edge label'=$X$] (c),
      (e) -- [boson, half left, edge label=$W$] (d),
    };
  \end{feynman}
    \end{tikzpicture}
        \begin{tikzpicture}
        \label{fig:loopC}
  \begin{feynman}[]
    \vertex (a) {$s$};
    \vertex [right=2.5cm of a] (b);
    \vertex [above right=2 cm of b] (f1) {$d$};
    \vertex [below right=2 cm of b] (c);
    \vertex [left=1.5 cm of b] (d);
    \vertex [left=0.5 cm of b] (e);
    \diagram* {
      (a) -- [fermion] (d) -- [fermion, edge label'= $u_i$] (e) -- [fermion, edge label'= $d$] (b) -- [fermion] (f1),
      (b) -- [boson, edge label'=$X$] (c),
      (e) -- [boson, half right, edge label'=$W$] (d),
    };
  \end{feynman}
    \end{tikzpicture}
    \caption{Feynman diagrams contributing to the process $K^\pm\to\pi^\pm X$ at the one-loop level.}
    \label{fig:OneLoopDiagrams}
\end{figure}
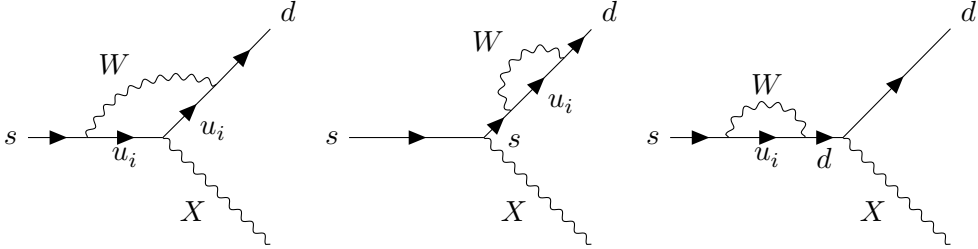

In this section, we will calculate the one-loop contributions to the flavour-violating process $K^\pm\to\pi^\pm X$. 
At the quark level, the Feynman diagrams generating the underlying 
$s\to d$ transition are displayed in Fig.~\ref{fig:OneLoopDiagrams}.
Notice that these diagrams are sensitive to different couplings 
of the $X$ boson to quarks. 
Summing up all contributions, the full amplitude reads
\begin{align}
    \mathcal{M} &= g_x x^{\rm{eff}}_{sd} \epsilon^*_\mu \bar{s}\gamma^\mu (1-\gamma_5) d = 
    \frac{g^2}{128\pi^2}\,g_x\,V_{id}V^*_{is} \, x_i \, \epsilon^*_\mu
    \bigg[x_R^{u_i}\left(\frac{2}{\epsilon}+\log{\frac{\mu^2}{m_i^2}}-\frac{1}{2}-3\frac{(1-x_i+\log{x_i})}{(x_i-1)^2}\right) \nonumber\\ 
    &+ \frac{(x_L^d+ x_L^s)}{2}\left(\frac{3-3x_i^2+2(4x_i-1)\log{x_i}}{2(x_i-1)^2}-\frac{2}{\epsilon} - \log\frac{\mu^2}{m_i^2}\right) + x_L^{u_i}\frac{(-1+x_i-4)}{(x_i-1)} \bigg] 
    \nonumber \\
&\times \bar{s}\gamma^\mu (1-\gamma_5) d \, , 
\end{align}
where $x_i=m_i^2/m_W^2$ with $m_i$ being the mass of the up-type quark running in the loop and we have defined the chiral charges as $x_{L/R}^f=x_V^f\mp x_A^f$. 
The divergences, originating from the non-renormalizability of 
our model, can be interpreted within a hard cutoff
regularization scheme as $2/\epsilon+\log(\mu^2/m_i^2)
=\log(\Lambda^2/m_i^2)$ where $\Lambda$ is the UV cutoff.
In specific renormalizable models, $\Lambda$ will be 
identified with the mass scale of particles belonging 
to the NP sector which will provide a UV completion of our model. 

In the limit of universal couplings, i.e.
$x_{V,A}^{u_i} = x_{V,A}^d = x_{V,A}^s$, 
and keeping only the dominant loop effects 
stemming from the exchange of the top quark,
we obtain
\begin{align}
\label{eq:OneLoopFlavourUniversal}
x^{\rm{eff}}_{sd} &\simeq 
\frac{g^2}{64\pi^2}\,V_{td}V^*_{ts} 
x_A^u f(x_t) 
\end{align}
where 
\begin{align}
f(x_t) = x_t
\left[\frac{2}{\epsilon}+\log{\frac{\mu^2}{m_t^2}}-\frac{1}{2}-3
\frac{(1-x_t+\log{x_t})}{(x_t-1)^2}\right] \, .
\end{align}
The inclusion of the above loop effects in the decay rate of 
$K^\pm\to\pi^\pm X$ can be implemented by the following replacement 
in Eq.~\eqref{amplitude_squared}:
\begin{equation}
\label{eq:shift}
x^{32}_V \to x^{32}_V - x^{\rm{eff}}_{sd}\,. 
\end{equation}
As a result, we can estimate the relative size of loop effects 
and tree-level ones as
\begin{align}
\label{eq:1_loop_vs_tree}
\frac{x^{\rm{eff}}_{sd}}{4 g_8 f_{\pi}^2 G x_A^u} \approx f(x_t) \, ,
\end{align}
where $f(x_t)$ is a model-dependent loop function which depends
on the specific UV completion of our effective theory. 
Therefore, we have learned that loop-effects have a similar size
of tree-level contributions. However, while the former suffer from sizeable uncertainties, the latter provide a robust model-independent result. Moreover, we also remark that loop and tree-level contributions 
generally depend on different couplings and therefore the comparison 
in Eq.~\eqref{eq:1_loop_vs_tree} is valid only in the universal scenario $x_{V,A}^{u_i} = x_{V,A}^d = x_{V,A}^s$.

\section{Flavour bounds vs.~beam-dump and collider searches}
\label{sec:numerics}

\begin{figure}[]
        \centering
        \includegraphics[width=0.7\textwidth]{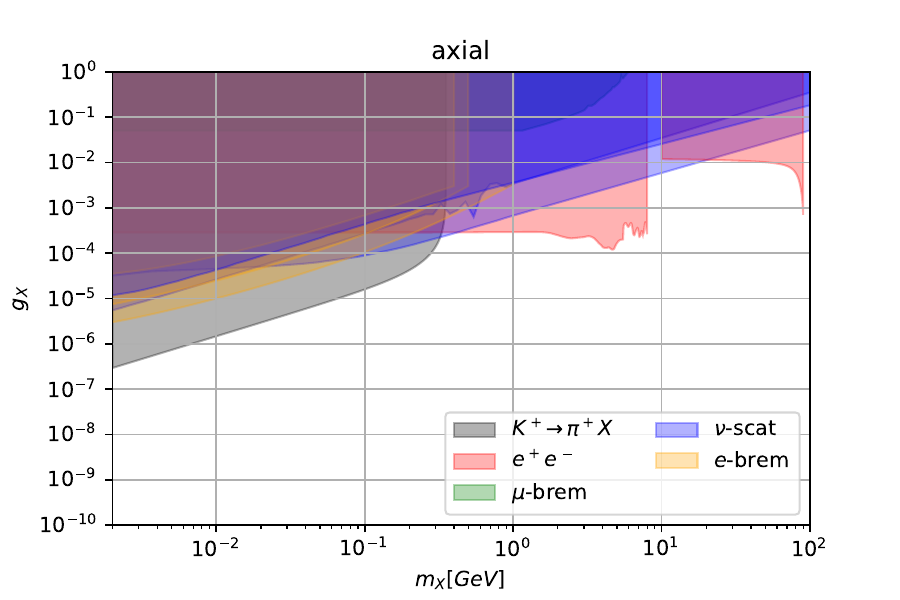}
        \includegraphics[width=0.7\textwidth]{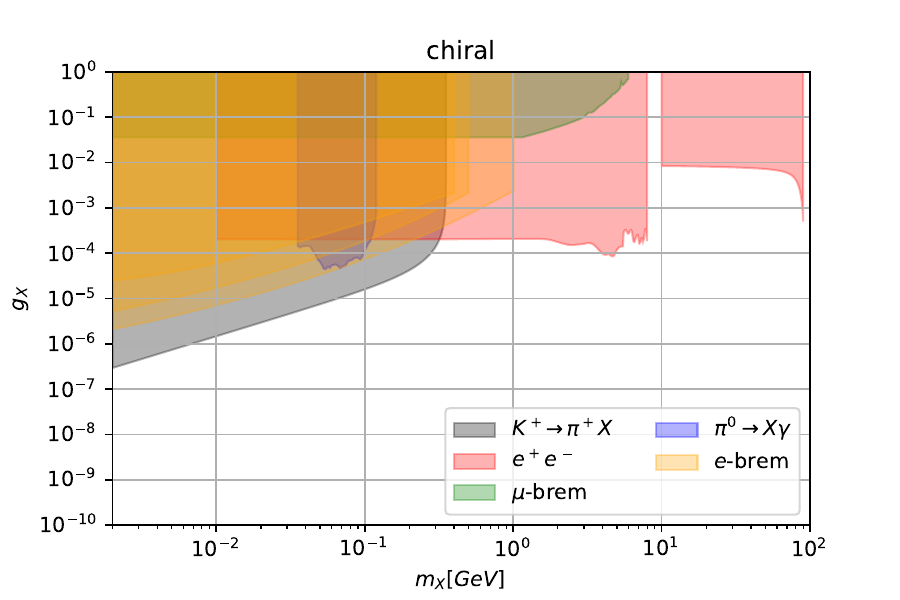} \\
        \includegraphics[width=0.7\textwidth]{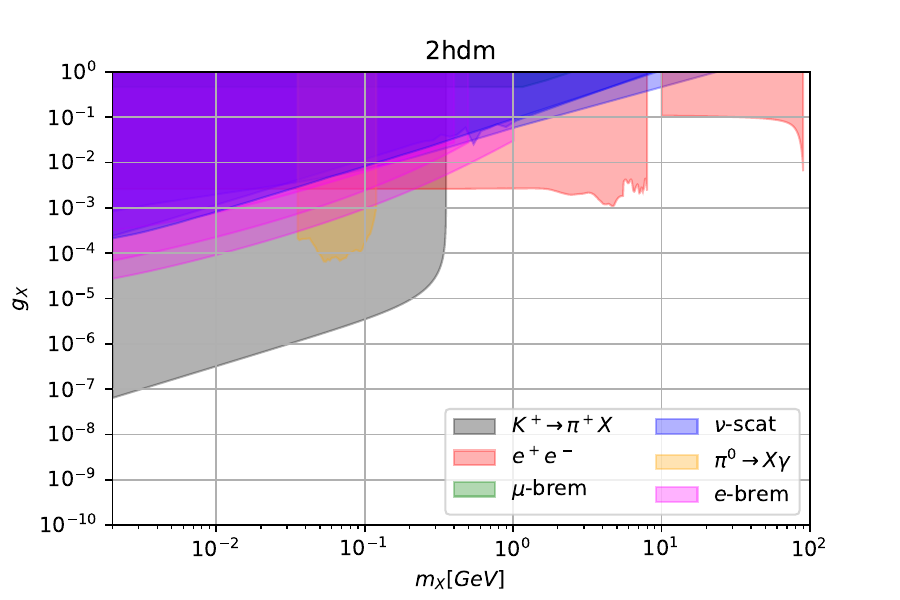} \\
        \caption{The dark shaded area represents the 
        tree-level $K^\pm\to\pi^\pm X$ bound obtained in this work. 
        Limits from beam-dump and collider searches are 
        obtained with DarkCast \cite{Baruch:2022esd} 
        and are shown for the purpose of comparison 
        for the three benchmark models 
        given in Table \ref{tab:model charges}.
        }
        \label{fig:Bounds}
    \end{figure}

    We are ready now to exploit the results derived in the previous section, 
    in order to explore the capability of the process $K^\pm\to\pi^\pm X$ 
    to unveil new light vector bosons.
    We are going to use the DarkCast package~\cite{Ilten:2018crw, Baruch:2022esd}, which enables us to derive bounds on vector and
    axial couplings of models entailing new spin-1 states by imposing
    current and future experimental constraints on several processes.
    In Fig.~\ref{fig:Bounds}, we show the bounds in the $(m_X,g_x)$ plane arising 
    from a variety of beam-dump and collider searches~\cite{Baruch:2022esd} as well 
    as from the flavour changing process $K^\pm\to\pi^\pm X$ discussed in this paper.  
    The three plots refer to the benchmark models dubbed as axial, chiral and 2HDM~\cite{Baruch:2022esd} which differ for the values of the $x_{V,A}$ charges, 
    see Table \ref{tab:model charges}. 
\begin{table}[ht]
    \centering
    \begin{tabular}{|c||c|c|c|c||c|c|c|c|}
        \hline
        & $x^e_V$ & $x^\nu_V$ & $x^{u,c,t}_V$ & $x^{d,s,b}_V$ & $x^e_A$ & $x^\nu_A$ & $x^{u,c,t}_A$ & $x^{d,s,b}_A$ \\
        \hline
        Axial & 0 & 1/4  & 0 & 0 & $-1$ & $-1/4$  & 1 & $-1$ \\
        Chiral & $-1$ & 0 & 1 & 1 & $-1$ & 0 & 1 & $-1$ \\
        2HDM & 0.044 & 0.05  & 1.021 & 0.015 & $-0.1$ & 0.05  & $-0.95$ & $-0.1$ \\
        \hline
    \end{tabular}
    \caption{Charges of the SM fermions under $X$ boson interactions for the models considered in Ref.~\cite{Baruch:2022esd} where, for simplicity, flavor universal couplings have been assumed.
    }
    \label{tab:model charges}
\end{table}

    The bounds from the process $K^\pm\to\pi^\pm X$ are obtained by employing the tree-level prediction of \sect{sec:tree}
and exploiting the 
measurement of $\text{BR}(K^+\to \pi^+\nu\nu)=( 1.73^{+1.15}_{-1.05})\times 10^{-10}$ by the E949 experiment at BNL~\cite{E949:2008btt}. In particular, we impose the $2\sigma$ bound $\text{BR}(K^+\to \pi^+X)\lesssim 4 \times 10^{-10}$.
    Remarkably, in all scenarios of Fig.~\ref{fig:Bounds}, the process $K^\pm\to\pi^\pm X$ 
    sets the strongest to date model-independent bound in the $(m_X,g_x)$ plane for $m_X < m_K - m_\pi$.  

\section{Conclusions}
\label{sec:concl}

Extensions of the SM entailing new feebly interacting massive particles with sub-GeV masses are currently among the most studied scenarios of NP. In particular, comprehensive analyses aiming to probe a light spin-1 boson $X$ featuring general couplings to quarks and leptons have been carried out in the literature exploiting 
beam-dump and collider searches.

In this work, we revisited the flavour constraints to this scenario
by means of the rare decay $K^\pm\to\pi^\pm X$. In particular, we extended previous studies by constructing the most general $\Delta S = 1$ chiral Lagrangian as induced by weak interactions up to the order $\mathcal{O}(p^4)$.

The lowest-order $\mathcal{O}(p^2)$ terms of our $\chi$PT capture the effects to $K^\pm\to\pi^\pm X$ discussed in~\cite{Dror:2017nsg}, which are loop-induced and suppressed by the fifth power of the Cabibbo angle. 
Instead, the inclusion of subleading $\mathcal{O}(p^4)$ terms in the chiral expansion generates $K^\pm\to\pi^\pm X$ already at the tree-level and the related amplitude is only singly Cabibbo suppressed.
%
As a result, rather surprisingly, the two contributions turn out to be 
of comparable size, see Eq.~\eqref{eq:1_loop_vs_tree}. 

However, while the tree-level weak effects discussed in this work are model-independent, the loop-induced contributions of Ref.~\cite{Dror:2017nsg} are instead sensitive to the specific UV completion accounting for the mass of the spin-1 boson.

In conclusion, we have shown that the process $K^\pm\to\pi^\pm X$ sets the strongest to date model-independent bound on the diagonal axial-vector couplings to $u,d,s$ quarks of a light $X$ with $m_X < m_K - m_\pi$, superseding the bounds arising from beam-dump and collider searches.

\section*{Acknowledgements}
 
This work received funding from the European Union's Horizon 2020 research and innovation programme under the 
Marie Sk\l{}odowska-Curie grant agreement N$^{\circ}$ 860881 and by the INFN Iniziative Specifica APINE. 
The work of LDL is also supported by the project ``CPV-Axion''
under the Supporting TAlent in ReSearch@University of Padova (STARS@UNIPD).


\appendix

\section{Details of the chiral Lagrangian construction} 
\label{appendix}

In this appendix we will provide an extensive derivation of 
the chiral Lagrangian describing $\Delta S=1$ transitions. 
The operators of Eq.~\eqref{eq:DeltaS1Operators} have the form
\begin{equation}
    \mathcal{L}^{\rm EW, \, LL}_{\rm eff} = [t_{L}]^{jl}_{ik} (\bar{q}_L^i \gamma^\mu q_{Lj}) (\bar{q}_L^k \gamma_\mu q_{Ll}) \, , 
    \label{eq:QuarkLagLL}
\end{equation}
and
\begin{equation}
    \mathcal{L}^{\rm EW, \, LR}_{\rm eff} = [t^{\delta \delta}_{LR}]^{jl}_{ik} (\bar{q}_L^i \gamma^\mu q_{Lj}) (\bar{q}_R^k \gamma_\mu q_{Rl}) + [t^{\lambda \lambda}_{LR}]^{jl}_{ik} (\bar{q}_L^i \gamma^\mu T^a q_{Lj}) (\bar{q}_R^k \gamma_\mu T^a q_{Rl}) \, .
    \label{eq:QuarkLagLR}
\end{equation}
%
Let us first discuss how to identify those combinations of four-quark operators belonging to irreducible representations of the chiral group~\cite{Pich:2021yll, RBC:2001pmy, Lehner:2011fz}. 
From the invariance under the flavour group $U(3)_F \equiv U(3)_L \times U(3)_R$, we know that 
the $LL$ currents transform as the 81-dimensional representation of $(\bar{3}\otimes3)\otimes (\bar{3}\otimes 3)$ of $U(3)_L$ which can be further decomposed into irreducible symmetric or antisymmetric representations having dimension $1, 8, 10, 27$. 


In particular, one has symmetric-symmetric combinations $\mathcal{S}\mathcal{S}$ transforming in the $1\oplus8\oplus27$ representations and  antisymmetric-antisymmetric combinations $\mathcal{A}\mathcal{A}$ transforming as $1\oplus8$ representations.
We can disregard the symmetric-antisymmetric and antisymmetric-symmetric combinations transforming in the $8 \oplus 10$ representations because they cannot be generated by the operators in \eqref{eq:QuarkLagLL} which 
are symmetric under the simultaneous exchange of upper indices with the lower ones, $(i,k) \leftrightarrow (j,l) $.
In any case, the particular combinations of quark currents transforming in each one of these representations can be obtained by projecting the fundamental structure in \eqref{eq:QuarkLagLL} on the orthonormal basis of irreducible representations of $(\bar{3}\otimes3)\otimes (\bar{3}\otimes 3)$:
\begin{equation}
   \{ (e^a_{27})^{\mathcal{S}}_{\mathcal{S}},\,  (e^a_8)^{\mathcal{S}}_{\mathcal{S}}, \, (e_1)^{\mathcal{S}}_{\mathcal{S}}, \, (e^a_{10})^{\mathcal{A}}_{\mathcal{S}}, \, (e^a_8)^{\mathcal{A}}_{\mathcal{S}}, \, (e^a_{10})^{\mathcal{S}}_{\mathcal{A}}, \, (e^a_8)^{\mathcal{S}}_{\mathcal{A}},\, (e^a_8)^{\mathcal{A}}_{\mathcal{A}}, \, (e_1)^{\mathcal{A}}_{\mathcal{A}} \} \, .
\end{equation}
This task is accomplished by making use of the tensorial product defined as
\begin{equation}
    \mathcal{T}_1\cdot \mathcal{T}_2 \equiv (T_1)^{ij}_{kl}(T_2)^{kl}_{ij}\,,\qquad\qquad
        \mathcal{T}^{a\,M}_{r\,N} \equiv \mathcal{T}\cdot e^{a \,M}_{r\,N} = T^{ij}_{kl} \, [e^{a \,M}_{r \, N}]^{kl}_{ij} \, , 
\end{equation}
where we exploited the decomposition $\mathcal{T} = \mathcal{T}^{a\,M}_{r \,N} e^{a \,M}_{r \,N}$.

Now, the fully symmetric singlet and octet basis elements are
\begin{equation}
        [(e_{1})_\mathcal{S}^{\mathcal{S}}]^{kl}_{ij} = \frac{1}{2\sqrt{6}} 
        [\delta^k_i \delta_j^l + \delta^l_i \delta_j^l]\,,\qquad 
        [(e^a_{8})_\mathcal{S}^{\mathcal{S}}]^{kl}_{ij} = \frac{1}{2\sqrt{10}} [(\lambda^a)^k_i\delta^l_j + (\lambda^a)^l_i\delta^k_j + (\lambda^a)^k_j\delta^l_i + (\lambda^a)^l_j\delta^k_i] \, , 
\end{equation}
whereas the fully antisymmetric singlet and octet basis elements are 
\begin{equation}
        [(e_{1})_\mathcal{A}^{\mathcal{A}}]^{kl}_{ij} = \frac{1}{2\sqrt{3}} [\delta^k_i \delta_j^l - \delta^l_i \delta_j^l]\,,\qquad
        [(e^a_{8})_\mathcal{A}^{\mathcal{A}}]^{kl}_{ij} = \frac{1}{2\sqrt{2}} [(\lambda^a)^k_i\delta^l_j - (\lambda^a)^l_i\delta^k_j - (\lambda^a)^k_j\delta^l_i + (\lambda^a)^l_j\delta^k_i] \, .
\end{equation}
The symmetric-symmetric 27-plet basis element is harder to construct and a better strategy is to extract the corresponding component by subtracting from a fully symmetric tensor its octet and singlet parts, namely
\begin{equation}
   (\mathcal{T}_{27})^{\mathcal{S}}_{\mathcal{S}} = \mathcal{T}^{\mathcal{S}}_{\mathcal{S}} - (\mathcal{T}_{8})^{\mathcal{S}}_{\mathcal{S}} - (\mathcal{T}_{1})^{\mathcal{S}}_{\mathcal{S}} = \mathcal{T}^{\mathcal{S}}_{\mathcal{S}} - [\mathcal{T}^{\mathcal{S}}_{\mathcal{S}}\cdot (e_1)^{\mathcal{S}}_{\mathcal{S}}] \,(e_1)^{\mathcal{S}}_{\mathcal{S}} - [\mathcal{T}^{\mathcal{S}}_{\mathcal{S}}\cdot (e^a_8)^{\mathcal{S}}_{\mathcal{S}}]\, (e^a_8)^{\mathcal{S}}_{\mathcal{S}} \, .
\end{equation}
The decomposition of the operators appearing in \eqref{eq:QuarkLagLL} can be easily performed by projecting the quark currents onto the orthonormal basis elements. Their chiral counterparts are then simply obtained by projecting the chiral equivalent of quark currents in equation \eqref{eq:ChiralEquivalent} onto the very same basis elements. 

{\bf{LL currents}:} the fully symmetric and anti-symmetric octet Lagrangian reads
\begin{align}
\label{eq:Lag8SS}
\begin{split}
\!\!\!\!\!\!    
\mathcal{L}_8^\mathcal{S(A)} =&~ a_8^\mathcal{S(A)} \frac{f_\pi^4}{80} 
    [t_L]^{jl}_{ik} 
    \left(\Tr{(\lambda^a L_\mu)} \Tr{L^\mu} \pm \Tr{(\lambda^a L_\mu L^\mu)}\right)\, \cdot \\
    &~[(\lambda^a)^k_i\delta^l_j \pm (\lambda^a)^l_i\delta^k_j \pm (\lambda^a)^k_j\delta^l_i + (\lambda^a)^l_j\delta^k_i]
\end{split}
\end{align}
In principle, one should consider also the structure
$\Tr{(\lambda^a(U^\dagger \chi + \chi^\dagger U))}$
along with $\Tr{(\lambda^a L^\mu L_\mu)}$.
However, these additional structures induce 
vacuum misalignment effects through Goldstone tadpoles and can be rotated away by properly
redefining the Goldstone fields~\cite{Pich:2021yll}.

The 27-plet Lagrangian term is finally given by 
\begin{align}
\label{eq:Lag27}
    \mathcal{L}_{27} & = a_{27}\frac{f_\pi^4}{8} 
    \{[t_L]^{jl}_{ik} \big ( [(L_\mu)^i_j (L^\mu)^k_l + (L_\mu)^k_j(L^\mu)^i_l]  - \frac{1}{12} \left( \Tr{(L_\mu L^\mu)} + \Tr{L_\mu} \Tr{L^\mu} \right) [\delta^i_j\delta^k_l + \delta^i_l \delta^k_j]
    \nonumber\\
    & -\frac{1}{10} \left(\Tr{(\lambda^a L_\mu L^\mu)} + \Tr{(\lambda^a L_\mu)} \Tr{L^\mu} \right)[(\lambda^a)^k_i\delta^l_j + (\lambda^a)^l_i\delta^k_j + (\lambda^a)^k_j\delta^l_i + (\lambda^a)^l_j\delta^k_i]
\big) \}\,.
\end{align}

In the expressions \eqref{eq:Lag8SS} and \eqref{eq:Lag27}, the parameters $a_8^{\mathcal{S}}$, $a_8^{\mathcal{A}}$ and $a_{27}$ parametrize our ignorance about the hadronization dynamics and are to be determined experimentally.

{\bf{LR currents}:} Since left-handed and right-handed currents transform under different $U(3)$ groups, these combinations correspond to the product of a $\bar{3} \oplus 3 = 1 \oplus 8$ representation in each chiral sector, resulting in four possible different structures transforming as $(1_L, 1_R)$, $(8_L, 1_R)$, $(1_L, 8_R)$ and $(8_L, 8_R)$.
We first identify the associated orthonormal basis
\begin{equation}
\!\!\!\!\!\!\!
    (e_{1_L,1_R})^{kl}_{ij} \!=\! \frac{\delta^j_i \delta^l_k}{3},~~ 
    (e^a_{8_L,1_R})^{kl}_{ij} \!=\! \frac{(\lambda^a_L)^j_i \delta^l_k}{\sqrt{6}},~~ (e^a_{1_L,8_R})^{kl}_{ij} \!=\! \frac{\delta^j_i (\lambda^b_R)^l_k}{\sqrt{6}},~~
    (e^{a,b}_{8_L,8_R})^{kl}_{ij} \!=\! \frac{(\lambda^a_L)^j_i (\lambda^b_R)^l_k}{2} \, , 
\end{equation}
which will be then used in order to project the appropriate structure onto the low-energy operators possessing definite chiral transformation properties. Exploiting the completeness relation 
\begin{equation}
    T^a_{ij}T^a_{kl} = \frac{1}{2}\delta_{il}\delta_{kj} - \frac{1}{2N_C} \delta_{ij}\delta_{lk} \, , 
\end{equation}
we can recast Eq.~\eqref{eq:QuarkLagLR} in the following form
\begin{equation}
    \mathcal{L}^{\rm EW, \, LR}_{\rm eff} = [t^{\delta \delta}_{LR}]^{jl}_{ik} (\bar{q}_L^i \gamma^\mu q_{Lj}) (\bar{q}_R^k \gamma_\mu q_{Rl}) - \frac{1}{2N_C}[t^{\lambda \lambda}_{LR}]^{jl}_{ik} (\bar{q}_L^i \gamma^\mu q_{Lj}) (\bar{q}_R^k \gamma_\mu q_{Rl}) \, . 
\end{equation}
At this point the Fierz identity 
\begin{equation}
    (\bar{q}_L^i \gamma^\mu q_{Lj}) (\bar{q}_R^k \gamma_\mu q_{Rl}) = -2 (\bar{q}_L^i q_{Rl}) (\bar{q}_R^k q_{Lj})
\end{equation}
can be used in order to identify the various operators.

The leading order chiral structure that is compatible with an $(8_L, 8_R)$ structure reads
\begin{equation}
    \mathcal{L}_{8_L, 8_R} = \frac{f_\pi^6}{4} \big( a^{\delta \delta}_{88}[t_{LR}^{\delta \delta}]^{jl}_{ik} + a^{\lambda \lambda}_{88}[t_{LR}^{\lambda \lambda}]^{jl}_{ik}\big) (\lambda^a_L)^{i}_j (\lambda^b_R)^{k}_l \Tr{(\lambda_L^a U^\dagger \lambda^b_R U)} + \mathcal{O} (p^2) \, .
\end{equation}
Instead, the structures transforming as $(8_L, 1_R)$ and $(1_L, 8_R)$ are
%
\begin{equation}
    \mathcal{L}_{8}^{LR} = \frac{f_\pi^4}{6} \big( a^{\delta \delta}_{LR}[t_{LR}^{\delta \delta}]^{jl}_{ik} + a^{\lambda \lambda}_{LR}[t_{LR}^{\lambda \lambda}]^{jl}_{ik}\big) \{\Tr{(\lambda_a L_\mu L^\mu)} (\lambda_L^a)^i_j \delta^k_l + \Tr{(\lambda_a R_\mu R^\mu)} \delta^i_j(\lambda_R^b)^k_l \}\,. 
\end{equation}
Also in this case the constants $a^{\delta \delta}_{88}$, $a^{\lambda \lambda}_{88}$, $a^{\delta \delta}_{LR}$ and $a^{\lambda \lambda}_{LR}$ parametrize our ignorance on the non-perturbative dynamics related to the hadronization process.

Combining all above results, one obtains the $\Delta S = 1$ chiral Lagrangian of Eq.~\eqref{eq:ChiptWeakSM} 
%
%
%
%
where the coefficients $g_{27}$, $g_8$ and $g_{\text{ew}}$ are functions of the non-perturbative effective parameters $a_{*}$ defined above, as well as of the Wilson coefficients entering Eqs.~\eqref{eq:QuarkLagLL} and \eqref{eq:QuarkLagLR}.

As shown in \cite{RBC:2001pmy,Lehner:2011fz}, the matching procedure requires to decompose the operators of Eq.~\eqref{eq:SMDelta1Lag} into operators having well-defined chiral transformation properties under the flavour group
\begin{equation}
\begin{split}
    &Q_1 = \frac{1}{10} Q_S^{(8,1), 1/2} + \frac{1}{15}Q_S^{(27,1), 1/2} + \frac{1}{3} Q_S^{(27,1), 3/2} + \frac{1}{2}Q_A^{(8,1), 1/2} \\
    &Q_2 = \frac{1}{10} Q_S^{(8,1), 1/2} + \frac{1}{15}Q_S^{(27,1), 1/2} + \frac{1}{3} Q_S^{(27,1), 3/2} - \frac{1}{2}Q_A^{(8,1), 1/2} \\
    &Q_3 = \frac{1}{2} Q_S^{(8,1), 1/2}  + \frac{1}{2}Q_A^{(8,1), 1/2} \\
    &\tilde{Q}_4 = \frac{1}{2} Q_S^{(8,1), 1/2}  - \frac{1}{2}Q_A^{(8,1), 1/2} \\
    &Q_9 = -\frac{1}{10} Q_S^{(8,1), 1/2} + \frac{1}{10}Q_S^{(27,1), 1/2} + \frac{1}{2} Q_S^{(27,1), 3/2} + \frac{1}{2}Q_A^{(8,1), 1/2} \\
    &\tilde{Q}_{10} = -\frac{1}{10} Q_S^{(8,1), 1/2} + \frac{1}{10}Q_S^{(27,1), 1/2} + \frac{1}{2} Q_S^{(27,1), 3/2} + \frac{1}{2}Q_A^{(8,1), 1/2} \, ,  \\
\end{split}
\end{equation}
where $1/2$ and $3/2$ in the superscripts denote the operator isospin properties, while $\tilde{Q}_4$ and $\tilde{Q}_{10}$ 
are the Fierzed counterparts of $Q_4$ and $Q_{10}$: 
\begin{equation}
    \tilde{Q}_4 = 4 \sum_q (\bar{s}_L \gamma^\mu q_L) (\bar{q}_L \gamma_\mu d_L) \quad \text{and} \quad \tilde{Q}_{10} = 6 \sum_q e_q (\bar{s}_L \gamma^\mu q_L) (\bar{q}_L \gamma_\mu d_L) \, . 
\end{equation}
Defining 
\begin{equation}
    Q^{(27,1)}_S = \frac{1}{9} \big[ Q_S^{(27,1), 1/2} + 5\, Q_S^{(27,1), 3/2} \big] \, , 
\end{equation}
one can then isolate in each operator the desired chiral structures transforming in the irreducible representations of the flavour group~\cite{Kambor:1989tz}.
Then, such structures can be directly translated into their $\chi$PT counterparts. 

As far as the $LR$ operators are concerned, we first need to recast the operators in Eq.~\eqref{eq:DeltaS1Operators} in a form compatible with Eq.~\eqref{eq:QuarkLagLR}. This is achieved by making use of the completeness relation
\begin{equation}
    \delta_{\beta \gamma} \delta_{\alpha \delta} = 2 T^a_{\alpha \gamma} T^a_{\beta \delta}+ \frac{1}{N_C} \delta_{\alpha \gamma}  \delta_{\beta \delta } \, , 
\end{equation}
which allows us to rewrite 
\begin{equation}
    \begin{split}
        (\bar{q}^i_\alpha \gamma^\mu \delta_{\beta \gamma}q^j_\gamma)(\bar{q}^k_\beta \gamma_\mu \delta_{\alpha \delta}q^l_\delta) &= 2(\bar{q}^i_\alpha \gamma^\mu T^a_{\alpha\gamma}q^j_\gamma)(\bar{q}^k_\beta \gamma_\mu T^a_{\beta \delta}q^l_\delta) + \frac{1}{N_C}(\bar{q}^i_\alpha \gamma^\mu \delta_{\alpha\gamma}q^j_\gamma)(\bar{q}^k_\beta \gamma_\mu \delta_{\beta \delta}q^l_\delta)
        \\
        & \equiv 2Q^{\lambda\lambda}_{ijkl} + \frac{1}{N_C}Q^{\delta \delta}_{ijkl} \, .
    \end{split}
\end{equation}
The chiral counterparts of the operators on the right-hand side are well known. Then, from 
\begin{equation}
    C_{5(7)} Q_{5(7)} + C_{6(8)} Q_{6(8)} = \bigg(C_{5(7)}+ \frac{C_{6(8)}}{N_C}\bigg) Q_{\delta \delta} + 2 C_{6(8)} Q_{\lambda\lambda}
\end{equation}
and 
\begin{equation}
    Q_5 = Q_5^{(8,1)} \qquad \text{and} \qquad Q_7 = \frac{1}{2}Q^{(8,8), 3/2}_{S} + \frac{1}{2}Q^{(8,8), 1/2}_{A} \, , 
\end{equation}
we can proceed with the matching procedure obtaining the following results:
%
%
\begin{eqnarray*}
        g_{27} &=& \frac{3}{5} a_{27}(\mu) \big(C_1 + C_2 + \frac{3}{2} C_9 + \frac{3}{2} C_{10}\big)(\mu)\\
        g_{8} &=& \frac{1}{10} a_{8}^{\mathcal{S}} (\mu) \big(C_1+C_2+5C_3+5C_4-C_9-C_{10}\big)(\mu) + \\
        &  & -\frac{1}{2} a_{8}^{\mathcal{A}} (\mu) \big(C_1-C_2+C_3-C_4+C_9-C_{10}\big)(\mu)  \\
        &  & + 4 \{ a_{LR}^{\delta \delta} (\mu)\big( C_5 + \frac{C_6}{N_C} \big)(\mu) + 2 a_{LR}^{\lambda \lambda}(\mu) C_6(\mu)\}\\
        g^S_{8} &=& \frac{1}{10} a_{8}^{\mathcal{S}} (\mu) \big(C_1+C_2+5C_3+5C_4-C_9-C_{10}\big)(\mu) + \\
        &  & +\frac{1}{2} a_{8}^{\mathcal{A}} (\mu) \big(C_1-C_2+C_3-C_4+C_9-C_{10}\big)(\mu)  \\
        e^2 g_8\, g_{\text{ew}} &=& 6 \{ a_{88}^{\delta \delta} (\mu)\big( C_7 + \frac{C_8}{N_C} \big)(\mu) + 2 a_{88}^{\lambda \lambda}(\mu) C_8(\mu)\}\,.
\end{eqnarray*}

\begin{small}

\bibliographystyle{utphys}
\bibliography{bibliography.bib}

\providecommand{\href}[2]{#2}\begingroup\raggedright\begin{thebibliography}{10}

\bibitem{Holdom:1985ag}
B.~Holdom, ``{Two U(1)'s and Epsilon Charge Shifts},''
  \href{http://dx.doi.org/10.1016/0370-2693(86)91377-8}{{\em Phys. Lett. B}
  {\bfseries 166} (1986) 196--198}.

\bibitem{Pospelov:2007mp}
M.~Pospelov, A.~Ritz, and M.~B. Voloshin, ``{Secluded WIMP Dark Matter},''
  \href{http://dx.doi.org/10.1016/j.physletb.2008.02.052}{{\em Phys. Lett. B}
  {\bfseries 662} (2008) 53--61},
  \href{http://arxiv.org/abs/0711.4866}{{\ttfamily arXiv:0711.4866 [hep-ph]}}.

\bibitem{Bjorken:2009mm}
J.~D. Bjorken, R.~Essig, P.~Schuster, and N.~Toro, ``{New Fixed-Target
  Experiments to Search for Dark Gauge Forces},''
  \href{http://dx.doi.org/10.1103/PhysRevD.80.075018}{{\em Phys. Rev. D}
  {\bfseries 80} (2009) 075018},
  \href{http://arxiv.org/abs/0906.0580}{{\ttfamily arXiv:0906.0580 [hep-ph]}}.

\bibitem{APEX:2011dww}
{\bfseries APEX} Collaboration, S.~Abrahamyan {\em et~al.}, ``{Search for a New
  Gauge Boson in Electron-Nucleus Fixed-Target Scattering by the APEX
  Experiment},'' \href{http://dx.doi.org/10.1103/PhysRevLett.107.191804}{{\em
  Phys. Rev. Lett.} {\bfseries 107} (2011) 191804},
  \href{http://arxiv.org/abs/1108.2750}{{\ttfamily arXiv:1108.2750 [hep-ex]}}.

\bibitem{Merkel:2014avp}
H.~Merkel {\em et~al.}, ``{Search at the Mainz Microtron for Light Massive
  Gauge Bosons Relevant for the Muon g-2 Anomaly},''
  \href{http://dx.doi.org/10.1103/PhysRevLett.112.221802}{{\em Phys. Rev.
  Lett.} {\bfseries 112} no.~22, (2014) 221802},
  \href{http://arxiv.org/abs/1404.5502}{{\ttfamily arXiv:1404.5502 [hep-ex]}}.

\bibitem{BaBar:2009lbr}
{\bfseries BaBar} Collaboration, B.~Aubert {\em et~al.}, ``{Search for Dimuon
  Decays of a Light Scalar Boson in Radiative Transitions Upsilon
  ---\ensuremath{>} gamma A0},''
  \href{http://dx.doi.org/10.1103/PhysRevLett.103.081803}{{\em Phys. Rev.
  Lett.} {\bfseries 103} (2009) 081803},
  \href{http://arxiv.org/abs/0905.4539}{{\ttfamily arXiv:0905.4539 [hep-ex]}}.

\bibitem{Curtin:2013fra}
D.~Curtin {\em et~al.}, ``{Exotic decays of the 125 GeV Higgs boson},''
  \href{http://dx.doi.org/10.1103/PhysRevD.90.075004}{{\em Phys. Rev. D}
  {\bfseries 90} no.~7, (2014) 075004},
  \href{http://arxiv.org/abs/1312.4992}{{\ttfamily arXiv:1312.4992 [hep-ph]}}.

\bibitem{BaBar:2014zli}
{\bfseries BaBar} Collaboration, J.~P. Lees {\em et~al.}, ``{Search for a Dark
  Photon in $e^+e^-$ Collisions at BaBar},''
  \href{http://dx.doi.org/10.1103/PhysRevLett.113.201801}{{\em Phys. Rev.
  Lett.} {\bfseries 113} no.~20, (2014) 201801},
  \href{http://arxiv.org/abs/1406.2980}{{\ttfamily arXiv:1406.2980 [hep-ex]}}.

\bibitem{LHCb:2017trq}
{\bfseries LHCb} Collaboration, R.~Aaij {\em et~al.}, ``{Search for Dark
  Photons Produced in 13 TeV $pp$ Collisions},''
  \href{http://dx.doi.org/10.1103/PhysRevLett.120.061801}{{\em Phys. Rev.
  Lett.} {\bfseries 120} no.~6, (2018) 061801},
  \href{http://arxiv.org/abs/1710.02867}{{\ttfamily arXiv:1710.02867
  [hep-ex]}}.

\bibitem{Anastasi:2015qla}
A.~Anastasi {\em et~al.}, ``{Limit on the production of a low-mass vector boson
  in $\mathrm{e}^{+}\mathrm{e}^{-} \to \mathrm{U}\gamma$, $\mathrm{U} \to
  \mathrm{e}^{+}\mathrm{e}^{-}$ with the KLOE experiment},''
  \href{http://dx.doi.org/10.1016/j.physletb.2015.10.003}{{\em Phys. Lett. B}
  {\bfseries 750} (2015) 633--637},
  \href{http://arxiv.org/abs/1509.00740}{{\ttfamily arXiv:1509.00740
  [hep-ex]}}.

\bibitem{LHCb:2019vmc}
{\bfseries LHCb} Collaboration, R.~Aaij {\em et~al.}, ``{Search for
  $A'\to\mu^+\mu^-$ Decays},''
  \href{http://dx.doi.org/10.1103/PhysRevLett.124.041801}{{\em Phys. Rev.
  Lett.} {\bfseries 124} no.~4, (2020) 041801},
  \href{http://arxiv.org/abs/1910.06926}{{\ttfamily arXiv:1910.06926
  [hep-ex]}}.

\bibitem{CMS:2019buh}
{\bfseries CMS} Collaboration, A.~M. Sirunyan {\em et~al.}, ``{Search for a
  Narrow Resonance Lighter than 200 GeV Decaying to a Pair of Muons in
  Proton-Proton Collisions at $\sqrt{s} =$ TeV},''
  \href{http://dx.doi.org/10.1103/PhysRevLett.124.131802}{{\em Phys. Rev.
  Lett.} {\bfseries 124} no.~13, (2020) 131802},
  \href{http://arxiv.org/abs/1912.04776}{{\ttfamily arXiv:1912.04776
  [hep-ex]}}.

\bibitem{Bernardi:1985ny}
G.~Bernardi {\em et~al.}, ``{Search for Neutrino Decay},''
  \href{http://dx.doi.org/10.1016/0370-2693(86)91602-3}{{\em Phys. Lett. B}
  {\bfseries 166} (1986) 479--483}.

\bibitem{KLOE-2:2011hhj}
{\bfseries KLOE-2} Collaboration, F.~Archilli {\em et~al.}, ``{Search for a
  vector gauge boson in $\phi$ meson decays with the KLOE detector},''
  \href{http://dx.doi.org/10.1016/j.physletb.2011.11.033}{{\em Phys. Lett. B}
  {\bfseries 706} (2012) 251--255},
  \href{http://arxiv.org/abs/1110.0411}{{\ttfamily arXiv:1110.0411 [hep-ex]}}.

\bibitem{KLOE-2:2012lii}
{\bfseries KLOE-2} Collaboration, D.~Babusci {\em et~al.}, ``{Limit on the
  production of a light vector gauge boson in phi meson decays with the KLOE
  detector},'' \href{http://dx.doi.org/10.1016/j.physletb.2013.01.067}{{\em
  Phys. Lett. B} {\bfseries 720} (2013) 111--115},
  \href{http://arxiv.org/abs/1210.3927}{{\ttfamily arXiv:1210.3927 [hep-ex]}}.

\bibitem{WASA-at-COSY:2013zom}
{\bfseries WASA-at-COSY} Collaboration, P.~Adlarson {\em et~al.}, ``{Search for
  a dark photon in the $\pi^0 \to e^+e^-\gamma$ decay},''
  \href{http://dx.doi.org/10.1016/j.physletb.2013.08.055}{{\em Phys. Lett. B}
  {\bfseries 726} (2013) 187--193},
  \href{http://arxiv.org/abs/1304.0671}{{\ttfamily arXiv:1304.0671 [hep-ex]}}.

\bibitem{HADES:2013nab}
{\bfseries HADES} Collaboration, G.~Agakishiev {\em et~al.}, ``{Searching a
  Dark Photon with HADES},''
  \href{http://dx.doi.org/10.1016/j.physletb.2014.02.035}{{\em Phys. Lett. B}
  {\bfseries 731} (2014) 265--271},
  \href{http://arxiv.org/abs/1311.0216}{{\ttfamily arXiv:1311.0216 [hep-ex]}}.

\bibitem{NA482:2015wmo}
{\bfseries NA48/2} Collaboration, J.~R. Batley {\em et~al.}, ``{Search for the
  dark photon in $\pi^0$ decays},''
  \href{http://dx.doi.org/10.1016/j.physletb.2015.04.068}{{\em Phys. Lett. B}
  {\bfseries 746} (2015) 178--185},
  \href{http://arxiv.org/abs/1504.00607}{{\ttfamily arXiv:1504.00607
  [hep-ex]}}.

\bibitem{NA62:2019meo}
{\bfseries NA62} Collaboration, E.~Cortina~Gil {\em et~al.}, ``{Search for
  production of an invisible dark photon in $\pi^0$ decays},''
  \href{http://dx.doi.org/10.1007/JHEP05(2019)182}{{\em JHEP} {\bfseries 05}
  (2019) 182}, \href{http://arxiv.org/abs/1903.08767}{{\ttfamily
  arXiv:1903.08767 [hep-ex]}}.

\bibitem{Kahn:2016vjr}
Y.~Kahn, G.~Krnjaic, S.~Mishra-Sharma, and T.~M.~P. Tait, ``{Light Weakly
  Coupled Axial Forces: Models, Constraints, and Projections},''
  \href{http://dx.doi.org/10.1007/JHEP05(2017)002}{{\em JHEP} {\bfseries 05}
  (2017) 002}, \href{http://arxiv.org/abs/1609.09072}{{\ttfamily
  arXiv:1609.09072 [hep-ph]}}.

\bibitem{Ilten:2018crw}
P.~Ilten, Y.~Soreq, M.~Williams, and W.~Xue, ``{Serendipity in dark photon
  searches},'' \href{http://dx.doi.org/10.1007/JHEP06(2018)004}{{\em JHEP}
  {\bfseries 06} (2018) 004}, \href{http://arxiv.org/abs/1801.04847}{{\ttfamily
  arXiv:1801.04847 [hep-ph]}}.

\bibitem{Baruch:2022esd}
C.~Baruch, P.~Ilten, Y.~Soreq, and M.~Williams, ``{Axial vectors in
  DarkCast},'' \href{http://arxiv.org/abs/2206.08563}{{\ttfamily
  arXiv:2206.08563 [hep-ph]}}.

\bibitem{Asai:2022zxw}
K.~Asai, A.~Das, J.~Li, T.~Nomura, and O.~Seto, ``{Chiral Z' in FASER, FASER2,
  DUNE, and ILC beam dump experiments},''
  \href{http://dx.doi.org/10.1103/PhysRevD.106.095033}{{\em Phys. Rev. D}
  {\bfseries 106} no.~9, (2022) 095033},
  \href{http://arxiv.org/abs/2206.12676}{{\ttfamily arXiv:2206.12676
  [hep-ph]}}.

\bibitem{Dror:2017ehi}
J.~A. Dror, R.~Lasenby, and M.~Pospelov, ``{New constraints on light vectors
  coupled to anomalous currents},''
  \href{http://dx.doi.org/10.1103/PhysRevLett.119.141803}{{\em Phys. Rev.
  Lett.} {\bfseries 119} no.~14, (2017) 141803},
  \href{http://arxiv.org/abs/1705.06726}{{\ttfamily arXiv:1705.06726
  [hep-ph]}}.

\bibitem{Dror:2017nsg}
J.~A. Dror, R.~Lasenby, and M.~Pospelov, ``{Dark forces coupled to nonconserved
  currents},'' \href{http://dx.doi.org/10.1103/PhysRevD.96.075036}{{\em Phys.
  Rev. D} {\bfseries 96} no.~7, (2017) 075036},
  \href{http://arxiv.org/abs/1707.01503}{{\ttfamily arXiv:1707.01503
  [hep-ph]}}.

\bibitem{Buchalla:1995vs}
G.~Buchalla, A.~J. Buras, and M.~E. Lautenbacher, ``{Weak decays beyond leading
  logarithms},'' \href{http://dx.doi.org/10.1103/RevModPhys.68.1125}{{\em Rev.
  Mod. Phys.} {\bfseries 68} (1996) 1125--1144},
  \href{http://arxiv.org/abs/hep-ph/9512380}{{\ttfamily arXiv:hep-ph/9512380}}.

\bibitem{DiLuzio:2022ziu}
L.~Di~Luzio, M.~Nardecchia, and C.~Toni, ``{Light vectors coupled to anomalous
  currents with harmless Wess-Zumino terms},''
  \href{http://dx.doi.org/10.1103/PhysRevD.105.115042}{{\em Phys. Rev. D}
  {\bfseries 105} no.~11, (2022) 115042},
  \href{http://arxiv.org/abs/2204.05945}{{\ttfamily arXiv:2204.05945
  [hep-ph]}}.

\bibitem{Gasser:1984gg}
J.~Gasser and H.~Leutwyler, ``{Chiral Perturbation Theory: Expansions in the
  Mass of the Strange Quark},''
  \href{http://dx.doi.org/10.1016/0550-3213(85)90492-4}{{\em Nucl. Phys. B}
  {\bfseries 250} (1985) 465--516}.

\bibitem{Pich:1995bw}
A.~Pich, ``{Chiral perturbation theory},''
  \href{http://dx.doi.org/10.1088/0034-4885/58/6/001}{{\em Rept. Prog. Phys.}
  {\bfseries 58} (1995) 563--610},
  \href{http://arxiv.org/abs/hep-ph/9502366}{{\ttfamily arXiv:hep-ph/9502366}}.

\bibitem{Pich:2021yll}
A.~Pich and A.~Rodr\'\i{}guez-S\'anchez, ``{SU(3) analysis of four-quark
  operators: $K\to\pi\pi$ and vacuum matrix elements},''
  \href{http://dx.doi.org/10.1007/JHEP06(2021)005}{{\em JHEP} {\bfseries 06}
  (2021) 005}, \href{http://arxiv.org/abs/2102.09308}{{\ttfamily
  arXiv:2102.09308 [hep-ph]}}.

\bibitem{RBC:2001pmy}
{\bfseries RBC} Collaboration, T.~Blum {\em et~al.}, ``{Kaon matrix elements
  and CP violation from quenched lattice QCD: 1. The three flavor case},''
  \href{http://dx.doi.org/10.1103/PhysRevD.68.114506}{{\em Phys. Rev. D}
  {\bfseries 68} (2003) 114506},
  \href{http://arxiv.org/abs/hep-lat/0110075}{{\ttfamily
  arXiv:hep-lat/0110075}}.

\bibitem{Lehner:2011fz}
C.~Lehner and C.~Sturm, ``{Matching factors for Delta S=1 four-quark operators
  in RI/SMOM schemes},''
  \href{http://dx.doi.org/10.1103/PhysRevD.84.014001}{{\em Phys. Rev. D}
  {\bfseries 84} (2011) 014001},
  \href{http://arxiv.org/abs/1104.4948}{{\ttfamily arXiv:1104.4948 [hep-ph]}}.

\bibitem{Gerard:2005yk}
J.-M. Gerard, C.~Smith, and S.~Trine, ``{Radiative kaon decays and the penguin
  contribution to the Delta I = 1/2 rule},''
  \href{http://dx.doi.org/10.1016/j.nuclphysb.2005.09.040}{{\em Nucl. Phys. B}
  {\bfseries 730} (2005) 1--36},
  \href{http://arxiv.org/abs/hep-ph/0508189}{{\ttfamily arXiv:hep-ph/0508189}}.

\bibitem{Ecker:2000zr}
G.~Ecker, G.~Isidori, G.~Muller, H.~Neufeld, and A.~Pich, ``{Electromagnetism
  in nonleptonic weak interactions},''
  \href{http://dx.doi.org/10.1016/S0550-3213(00)00568-X}{{\em Nucl. Phys. B}
  {\bfseries 591} (2000) 419--434},
  \href{http://arxiv.org/abs/hep-ph/0006172}{{\ttfamily arXiv:hep-ph/0006172}}.

\bibitem{E949:2008btt}
{\bfseries E949} Collaboration, A.~V. Artamonov {\em et~al.}, ``{New
  measurement of the $K^{+} \to \pi^{+} \nu \bar{\nu}$ branching ratio},''
  \href{http://dx.doi.org/10.1103/PhysRevLett.101.191802}{{\em Phys. Rev.
  Lett.} {\bfseries 101} (2008) 191802},
  \href{http://arxiv.org/abs/0808.2459}{{\ttfamily arXiv:0808.2459 [hep-ex]}}.

\bibitem{Kambor:1989tz}
J.~Kambor, J.~H. Missimer, and D.~Wyler, ``{The Chiral Loop Expansion of the
  Nonleptonic Weak Interactions of Mesons},''
  \href{http://dx.doi.org/10.1016/0550-3213(90)90236-7}{{\em Nucl. Phys. B}
  {\bfseries 346} (1990) 17--64}.

\end{thebibliography}\endgroup

\end{small}

\clearpage 

\end{document}